\documentclass[aip,]{revtex4-1}
\usepackage{graphicx}
\usepackage{dcolumn}
\usepackage{bm}
\usepackage{amsmath}
\usepackage[latin1]{inputenc}
\usepackage[frenchb]{babel}

\begin{document}


\title{Automatic detection and tracking of dust particles in a RF plasma sheath} 



\author{Y. Zayachuk}
\author{F. Brochard}
\email[]{frederic.brochard@ijl.nancy-universite.fr}
\author{S. Bardin}
\author{J-L. Briançon}
\author{R. Hugon}
\author{J. Bougdira}
\affiliation{Institut Jean Lamour, UMR 7198 CNRS, Nancy-Université, Faculté des Sciences et Techniques, entrée 3A, Boulevard des Aiguillettes, BP 70239, F-54506 Vandoeuvre-lès-Nancy Cedex, France}


\date{\today}

\begin{abstract}
A method enabling automatic detection and tracking of large amounts of individual dust particles in plasmas is presented. Individual trajectories can be found with a good spatiotemporal resolution, even without applying any external light source to facilitate detection. Main advantages of this method is a high portability and the possibility of making statistical analyses of the trajectories of a large amount of non uniformly size distributed particles, under challenging illumination conditions and with low time and computational resources. In order to evaluate the efficiency of different detection and tracking strategies statistically, an experimental benchmark is proposed, and completed by numerical simulations. 
\end{abstract}


\maketitle 

\section{Introduction}
Detection and tracking of punctual objects in order to analyze their individual displacement is a challenging issue that can meet a broad audience in the scientific community. As a few examples, one can cite from literature applications to cell tracking in medical imagery \cite{sbalza}, tracking in colloids \cite{crocker}, insect tracking \cite{lush}, or tracking of dust particles in fusion plasmas \cite{roque}. In these domains, the knowledge of individual object trajectories is needed, and techniques like Particle Image Velocimetry (PIV) \cite{grue}, developed to characterize fluid motion from the global displacement of fluid tracers are not relevant. According to the goal of the study, the method of investigation may differ. Our main goal is to develop a method which makes it possible to automatically detect and track a large number of individual dust particles in plasmas. However, with the proper set of parameters, the method proposed in this paper can be applied to a large variety of domains, where the detection and tracking of individual small object is required.

The term plasma dust stands for mesoscopic (size of nanometers to millimeters) charged particles, contained in plasmas. Dust is present in smaller or larger amounts in all plasmas used for practical purposes - both cold technological and hot fusion ones. There are two mechanisms responsible for that. First of them is sputtering - dust particles are formed due to the agglomeration of atoms and ions sputtered from the material wall of the device during the discharge. In fusion devices, UFOs and drops of melted armor material, floating in plasma  and formed during transient event such as disruption or Edge Localized Modes (ELMs) are considered to be dust as well \cite{winter}. Another possible dust formation process is agglomeration of ions and radicals directly in plasma volume, which takes place in cold technological plasmas \cite{hong}. 

In most cases the presence of dust is a limiting factor. In cold plasmas used for surface processing or thin films growth, deposition of dust on surfaces being processed leads to contamination of those surfaces resulting in degradation of their properties \cite{berndt}. In fusion plasmas several problems arise due to dust \cite{winter,hong,berndt,feder}. First of all, dust particles impinge on first wall of fusion device, where they can cause significant mechanical damage, leading to enhanced first wall erosion. In fusion devices using carbon as part of the first wall or divertor material, the retention of hydrogen isotopes by carbonaceous dust particles may lead to significant fusion fuel loss \cite{arnasjnm, kreter}. Tritium retention by dust deposited on the walls leads to enhanced radioactive contamination of the device itself \cite{kreterprl} and large quantities of dust in the vacuum vessel could be vulnerable to ignition during accidental ingress of air or water and transport radioactive materials in the case of an accident. If a substantial amount of dust reach the fusion core, it becomes a significant source of impurities that can decrease the energy output of the device \cite{krash}. Finally, dust deposition itself is undesirable, because deposited dust particles may fill slits or apertures created for engineering purposes, and also lower the efficiency of actively cooled Plasma Facing Components (PFCs) \cite{magaud}.

Consequently, it is necessary to find some way to eliminate plasma dust. There can be two possible approaches. The first one is to find a way to minimize or suppress dust generation and growth. The second one is to remove dust particles from the region of device where their presence is the most unwanted and transport them to a safer place, which requires investigation of details of dust transport. 

References \cite{vaulina, fortov1, fortov2, arnaspre, wang, roque} reports about some earlier experiments performed in the field of experimental research of plasma dust transport. All those experiments share some common features. The tracked particles are most often artificially produced, so that their size and mass is well defined, and injected into plasma. Their amount is usually small (several up to several dozens). For particles to be observable, dust cloud is usually highlighted by external light source. One possible technique of dust motion analysis is Particle Image Velocimetry (PIV) \cite{grue, brochard}, commonly used in the fluids community to visualize fluid motion by looking at passive tracers injected for that purpose in the fluid medium. However, PIV allows only investigating behaviour of dust system as a whole, and not one of single particles which may interact with each other. In plasma physics, individual dust behaviour was studied by superposing several subsequent frames, so that superposed images of some chosen dust particles produce trajectory of this particle \cite{vaulina}. Tracking algorithm are also used, but existing algorithms require working in ideal observation conditions enabled by the use of external light sources. When using lasers, great care should then be taken in order to avoid perturbations due to particle heating by the laser beam \cite{liu1, liu2}. 

There exist significant differences between those earlier studies and the one reported in this paper. In our case dust particles were generated in situ and not injected from outside the device. Amount of dust produced in this way may be large \cite{brochard, peng}. External light source was not applied. On the contrary, dust particles were observed as dark spots at bright - due to plasma luminescence - background. Dust trajectories were automatically found using a dedicated algorithm, able to detect dust particles at each consecutive frame and to track individual trajectories, giving as output temporal dependence of particles coordinates. Experimental setup used in our work is less favorable than the one usually applied but its configuration is closer to those met in numerous practical applications. Thus, main goal of this paper is to present a powerful technique of automatic particle detection and tracking, which could be rather easily implemented in different configurations and experimental conditions for investigating plasma dust.

The rest of the article is organized as follows: in  Sec.~II we describe the experimental setup. Methods used for data analysis, i.e. dust detection and tracking are presented in Sec.~III. The efficiency of the different methods and the validation of the results is a challenging and central point, which is discussed in Sec.~IV. In this section, we present how our experimental arrangement can be adapted to benchmark our algorithm in real experimental conditions. Finally, key findings and prospects are summarized in Sec. V.
%
%
\section{Experimental set-up}
\begin{figure}
 \includegraphics[width=0.5\textwidth]{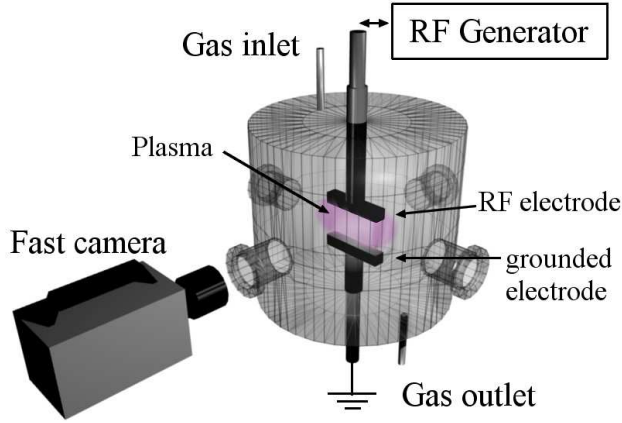}%
 \caption{\label{Setup} Scheme of the experimental set-up}%
 \end{figure}

Dust particles were generated in a capacitively coupled RF parallel plate reactor working at frequency 13.56 MHz, with upper electrode being biased and lower one grounded. Both electrodes are made of stainless steel. The experiment is depicted in Fig. \ref{Setup}. Vacuum in reactor chamber of this device is created by means of rotary vane pump. Best vacuum which can be achieved is 3 mTorr. Plasma is initiated in a mixture of argon and acetylene (C$_{2}$H$_{2}$). The use of hydrocarbon gas is necessary to provide carbon radicals used by particles to grow, since stainless steel electrodes are used. Graphite electrode can also be used, but dust particle produced in that way are smaller \cite{jqsrt}. The motion of dust particles is recorded by an ultrafast camera. Besides direct detection and tracking of dust particles, camera observation allows obtaining information about sheath dynamics, such as direct measurements of sheath luminescence intensity.

Design of plasma generator is such that it allows making discharges in gas mixes. Gases are continuously pumped out, simultaneously with gas injection. Mass flow meters control gas injection, which is set independently for each component of desired gas mixture, defining the composition of plasma. Usually electrodes in RF-generators of this kind are flat and circular. However, for experiments performed in this study it is necessary for dust particles to move approximately in single vertical plane (more detailed explanations are given in Sec. III). For this reason, electrodes used in this particular device were rectangular, with length 7 cm, width 2 cm and thickness 0.7 cm. Power is transmitted from generator to electrodes via matching box.
The key element of experimental setup is ultrafast camera FASTCAM SA1.1 by Photron Ltd, equipped with Nikon standard F-mount. Image recording is provided by a 12 bits monochrome CMOS sensor, with a spatial resolution of one megapixel up to 5400 fps. To improve resolution, external objective Nikon Micro-NIKKOR 105 mm with extension ring or Micro-NIKKOR 200 mm were mounted, providing a resolution of 15 $µm$ per pixel.

Experiments are conducted in the following way. Initially chamber is filled with pure argon. Then, discharge is initiated in argon at RF power 20 W. This power is being applied for $\sim$1 min, until discharge is stabilized. After that, power is set to 10 W, otherwise dust particles do not grow to size, large enough for purposes of experiment. Next setp consists in injecting acetylene. The acetylene fraction is set to 14\% and total pressure of gas mixture is 80 mTorr. These values are defined by the requirements of experiment: if acetylene fraction or pressure is lower, dust particles do not grow to sufficient size. After a few minutes, large ($\sim$100 $\mu$m) dust particle appear. Video recording using fast camera starts. Frame rate providing the clearest image is 500 fps. For this framerate, particle mean displacement between two frames is about 1 pixel, which is well suited for tracking. Finally, after recording, RF power is switched off and gas injection is stopped.

%
%
\section{Particle detection and tracking}

\subsection{Image processing}

As it is explained in next section, algorithm looks for pixels with brightness exceeding a threshold value, i.e. for bright spots on dark background. Recorded frames, however, contain bright background (which is in fact plasma luminescence) and dark dust particles. Thus, first operation to be done is to invert colormap of presented camera measurements.

\begin{figure}
 \includegraphics[width=0.5\textwidth]{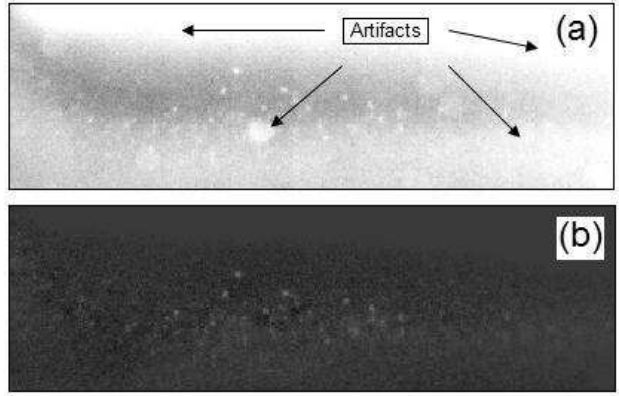}%
 \caption{\label{arti} Removal of artifacts: (a) initial 'dirty' frame (in inverted colors), (b) the same image with background removal.}%
 \end{figure}

In all experimentally recorded movies are present some "artifacts" - large motionless details, which are dirt on camera objective or observation window, or construction details, e.g., cathode. Since brightness of corresponding pixels may be large (in our case, some are brighter than dust particles), they may be recognized by software as possible particles, producing large number of misdetections, and making particle tracking much more difficult. A simple way to remove artifacts consists in subtracting averaged background to each frame (e.g., Fig. \ref{arti}). However, this procedure has no effect on fluctuating structures, such as those resulting from plasma fluctuations. In order to minimize misdetection resulting from such fluctuations, two possibilities preserving time and computational resources exist: if framerate is large compared to the timescale of fluctuations, a simple way to remove them is to remove a "gliding" background, consiting for frame $F_{i}$ of the average frame $(F_{i-1}+F_{i+1})/2$. In the other case, satisfying results can be obtained by performing a 2D high-pass FFT filtering on each individual frame. This option is the one retained in this paper when fixed threshold is used.

\subsection{Principle of the algorithm}
\subsubsection{Particle detection}
\begin{figure}
 \includegraphics[width=0.5\textwidth]{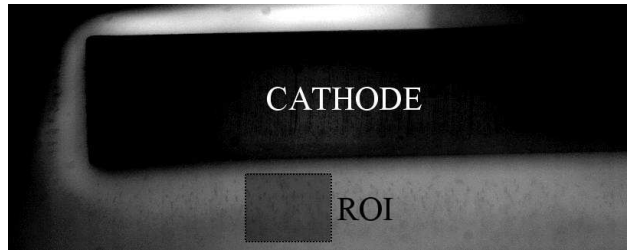}%
 \caption{\label{ROI} Camera frame, with contour showing the region of interest (ROI) used for the investigations presented in this paper.}%
 \end{figure}

Due to plasma luminescence, in our camera measurements dust particles appear as dark spot on a bright background (cf. Fig.~\ref{ROI}). In fusion plasmas, dust particles are easily seen as bright spots on a dark background, and in biology, cells are evidenced due to their sharp contour on a clear background. In all cases, the human brain performs contrast computation to evidence individual particles. Our algorithm does the same. More precisely, particle detection is based on a two-step procedure: image segmentation by intensity thresholding, and local intensity peak finding. 

First step, image segmentation, aims at simplifying the information contained in the image, by separating the image foreground (which contains the particles of interest) from the image background (everything else). This is done by comparing the value of each pixel with a threshold $T_{hr}$. Every pixel with a brightness intensity larger than $T_{hr}$ is considered as belonging to the foreground, otherwise it is considered as a part of the background, and set equal to zero. There are several ways to choose the value of $T_{hr}$. It can be fixed and imposed by the user, or computed automatically according to different techniques \cite{sezgin}. Moreover, it can be global (same $T_{hr}$ for the whole image) or locally adaptive (i.e., adapted to the different regions or scales of interest). In this paper fixed, automatic global and locally adaptive thresholds are used (refer to section~\ref{param} for more details). In particular cases, it can be interesting to use a combination of several techniques. For instance, the application to our algorithm of a hybrid technique mixing automatic global and locally adaptive threshold is prensented in Ref.\cite{ends}, in order to successfully process noisy data recorded during disruptions in a fusion plasma.

Second step is mainly devoted to the identification of particles coordinates.  Pixels forming the foreground are examined according to their brightness, either in ascending or descending order. In ascending order, pixels with intensity just above $T_{hr}$ are considered first. They are compared to their 8 direct neighbours. If none of the neighbours is brighter, pixel is a local maximum of intensity. In the opposite case, procedure is re-iterated around the brightest neighbour until a local maximum is found. When all the pixels of the foreground have been examined, a further step is needed to determine whether close local maxima correspond to individual particles or are part of the same particles. The latter case may be encountered when rather large particles with inhomogeneous brightness are present. Particle discrimination requires the knowledge of maximum particle radius $R_{p}$. All local maxima are examined (again either in ascending or descending order) and compared to any other local maximum located within a radius $r \leq R_{p}$ around it. In presented results, we assumed that the coordinates of a particle is given by the coordinates of its brightest pixel. This simplification is acceptable in our case where a 1-pixel accuracy is enough (compared to characteristic particle size and trajectories). If a better resolution is needed, sub-pixel resolution can be achieved by using weighted centroid approximation. In present paper, descending order has been chosen. We would like to point out that choosing ascending and descending orders may yield substantially different results.


\subsubsection{Particle tracking}
Particle detection algorithm gives as output arrays of particle coordinates for each frame. The disadvantage of method is that it produces no logical connection between particles detected at frame T and ones, detected at frame T+1, and it is unknown which particle at frame T corresponds to given particle at frame T+1. To perform tracking, it is necessary to specify one additional parameter, i.e. the maximal expected displacement between two consecutive frames, $D_{M}$. 

Our algorithm enables four different tracking strategies (TS) to operate trajectory reconstruction:
\subparagraph{TS-1. Minimization of Total Squared Displacement (TSD).}
This is the former approach used by Crocker and Grier \cite{crocker}. Having the set of  detected particles $N_{P}$ for frame T, code looks for all detected particles at frame T+1 which are located within the range $D_{M}$ from each of them. If in this range several particles are found, code calculates total squared displacement for all possible combinations of particle moves. Combination providing minimal total squared displacement $\sum_{i=1}^{N_{P}} \sqrt{(\Delta x_{i})^{2} + (\Delta y_{i})^{2}}$ is chosen as the most probable connection between particles. This strategy is very efficient when only a few particles are found within $D_{M}$, i.e. when the density of particulate medium is not to large and/or when framing rate is large with respect to particle displacement. It is the strategy used by default.

\subparagraph{TS-2. Oriented TSD approach (OTSD).}
This method is a refinement of the previous one. Instead of examining all the possible combinations of particle moves and keeping the one minimizing TSD, possible combinations are examined in a predefined order. Indeed, there are usually only a few allowed displacements for particles located at the edge of dense areas, where a large number of particles are found within $D_{M}$. Hence, this TS processes edge displacements first, which permits to lower the total number of combinations to be considered. This approach is more robust and much faster than the previous one when the particle density is very large.

\subparagraph{TS-3. Brightness Dependant TSD.}
In some experimental conditions, particles can be discriminated according to their brightness $B$ (e.g. dust particle in fusion plasmas \cite{ends}). In such conditions, taking into account particle brightness can be very efficient to improve the efficiency of particle tracking. In this case, particle brightness is included as a complementary parameter, in addition to space coordinates, in the computation of TSD. Trajectories retained are those which minimize the quantity $\sum_{i=1}^{N_{P}} \sqrt{(\Delta x_{i})^{2} + (\Delta y_{i})^{2} + (\Delta B_{i})^{2}}$.

\subparagraph{TS-4. Displacement extrapolation.}
With this method, when several particles are found within the range $D_{M}$, trajectories are found by extrapolating previous known particle positions. This method seems attractive to avoid confusion due to particle crossing, however it has several drawbacks. First of all, depending on the number of previous steps to take into account, it may lead untolerable computation time, especially in dense particulate media. Second, results are very sensitive to the extrapolation method, which has to be chosen according to the proper dynamics of the particulate medium under investigation, and might be rather difficult to determine \textit{a priori}. Trajectory extrapolation is ordinary applied to passive tracers in laminar or turbulent flows \cite{ouel}. In our case, particles are all but passive tracers: on the contrary they tend to avoid each other due to repulsive coulombian force, and exhibit a rather stochastic motion. For these reasons, we do not use any extrapolation at all in the presented results.

\subsection{Main parameters of the algorithm}
\label{param}
Although once properly configured, the algorithm can adapt automatically threshold value for performing particle detection, several parameters have to be defined by the user in order to obtain correct tracking. These parameters intimately depend on the detection system and on the main characteristics of the particulate medium. We present here a non-exhaustive list of the main configurable parameters:
\subsubsection{Detection parameters}
- Threshold parameter: in simple cases (i.e. bright particles, good contrast and no light variations) it can be enough to use a constant threshold value $T_{hr}$. On the contrary, in case of light intensity variations or inhomogeneous illumination, automatic thresholding is mandatory to obtain satisfactory results. Using local thresholding, best results were obtained by taking the threshold value equal to the mean value of non-null pixels in a 7$\times$7 box centered on the pixel of interest. With global thresholding, best results were obtained by using $T_{hr,g}~=~M-S/2$ where M and S are respectively the mean value of non-zero pixels and the standard deviation of the entire frame.

- Background: can be removed or not. When a fixed threshold is used, removing background can significantly improve detection of moving objects. On the contrary, removing background usually does not improve detection efficiency when automatic threshold is applied.

- Image preprocessing: the image can be filtered out using 2D FFT in order to disregard spatial scales smaller than $L_{noise}$ or larger than $L_{Max}$.

- Noise value: when working with automatic thresholding and noisy data, it can be useful to specify a minimal intensity threshold $I_{noise}$ equal to the intensity value of noise determined by the user according to the technical specification of the camera sensor. By default, $I_{noise}$ is taken equal to $I_{Min}+(I_{Max}-I_{Min})/10$, where $I_{Min}$ and $I_{Max}$ are respectively the lowest and highest intensity values in a region of 32 pixels $\times$ 32 pixels around the pixel of interest.

- Local filter size: when working with local thresholding, it is necessary to give the characteristic length on which local threshold $T_{loc}$ is determined.

- Particle maximum radius $R_{p}$: if large particles are observed, one has to check whether close intensity maxima belong to the same particle or not. If the distance between 2 intensity maxima is less than $R_{p}$, these 2 maxima are considered to be part of the same particle.

\subsubsection{Tracking parameters}
- Maximum displacement $D_{M}$: this is the maximum displacement allowed between consecutive frames. Framing rate should be high enough to keep $D_{M}$ low, since computation time and ressources as well as the probability of confusion between particles increase drastically with this parameter. On the other hand, if one tracked particle moves more than $D_{M}$ between 2 frames, then it is considered as being two different particles.

- Maximum recovery time: $T_{rec}$. This is the amount of time, in frame number, a particle can be lost before being found again. Particles can be lost mainly because of occlusions, because they leave the focal plane of the camera, or because of inadequate threshold. In dense particulate media, high values of $T_{rec}$ are not suited since this parameter affects drastically computational ressources, and also because the probability of reconnecting different trajectory segments of the same particle rapidly decays with particle density. 

- Tracking strategy: to be chosen between the four possibilities detailed in Sec. III.B.2.

- Extrapolation: when TS-4 is chosen as tracking strategy, the extrapolation method has to be chosen between linear or spline. More sophisticated extrapolation methods are not implemented yet.
%
%
\section{Implementing and validating automatic detection and tracking}

\subsection{Manual checking}
\begin{figure}
 \includegraphics[width=0.5\textwidth]{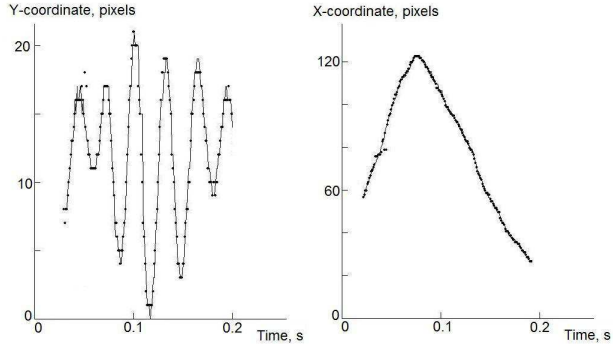}%
 \caption{\label{compmanual} Comparison between results of manual (firm line) and automatic tracking (dots), for one given particle. Y-coordinate stands for vertical coordinate, X-coordinate for horizontal one.}%
 \end{figure}

In order to check the tracking results, it is possible to compare the trajectories obtained automatically with manually tracked particles. This procedure is very long and fastidious and cannot be applied to a statistically relevant number of particles. Thus, we use manual checking mainly to test applicability and to have a first estimation of the accuracy of the algorithm. This procedure was applied to videos for which manual tracking of a few tens of particles was previously performed \cite{brochard}, and one example is given in Fig.~\ref{compmanual}. Deviation between manual and automatic trajectories turned out to be not larger than 1 pixel, i.e. about 20~$\mu$m. This precision is good enough for purposes of our experiment. It has to be mentioned that under difficult illumination conditions, manual tracking is extremely challenging and requires applying LUT corrections frame by frame. Under these conditions, 1 pixel accuracy is the very best that can be achieved manually.

%
%
\subsection{Experimental benchmark for statistical approach}
In the kind of RF plasma device used in our experiments, dust particle can exhibit a large variety of dynamics depending on their size, density, and on experimental conditions (gas flow, cathode bias, electrode and reactor geometry, pressure, instabilities etc...). Actually, characterizing the influence of some experimental parameters on dust transport was the main motivation for this work. In most cases, particle are characterized by fractionnal brownian motion, stochastic at short time scales and rather slightly superdiffusive at long time scales \cite{raty}, but antipersistent behaviour can also be observed \cite{brochard, bard}. In order to validate options used for automatic tracking, the ideal solution consists in comparing a statistical number of trajectories with reference trajectories. Practically, however, it is impossible to obtain statistical number of trajectories by hand. Thus we decided to influence dust behaviour by adapting our experimental set up, in order to compare the recorded trajectories with known main features of dust particle motion.

Dust particles in plasma sheath are subject to the action of various forces which confine them in the plasma volume or drag them outside. All the forces in competition depend on particle size, and can be distinguished in two classes: those independent of particle electric charge (gravitational force $F_{g}$, neutral drag force $F_{n}$ and thermophoretic force $F_{th}$) and those directly determined by the charge (electrostatic force $F_{el}$, ion drag force $F_{i}$, and repulsion force between particles $F_{dd}$) \cite{plain}. In our experimental conditions, large particles (15~$\mu$m in diameter or more) observed with the camera are mainly sensitive to $F_{g}$, $F_{n}$, and to all the forces depending on particle charge. Thus a simple way to influence dust displacement consist in perturbing the electric field in the plasma sheath. 

\begin{figure}
 \includegraphics[width=0.5\textwidth]{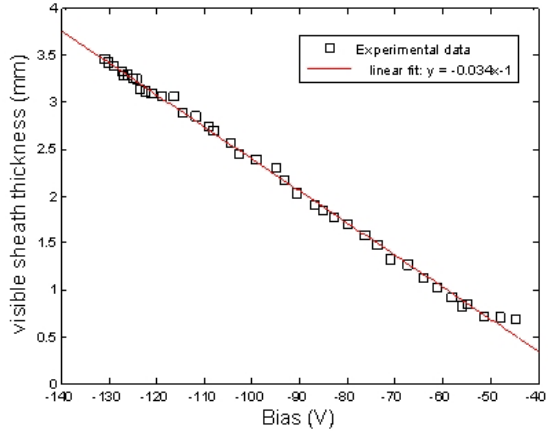}%
 \caption{\label{figsheath} Dependency of visible sheath thickness on cathode's self-bias (squares) and linear best fit (firm line). Sheath thickness is measured as the distance from the cathode to the maximum luminescence gradient. The error bars stand within the square symbols.}%
 \end{figure}

\begin{figure}
 \includegraphics[width=0.75\textwidth]{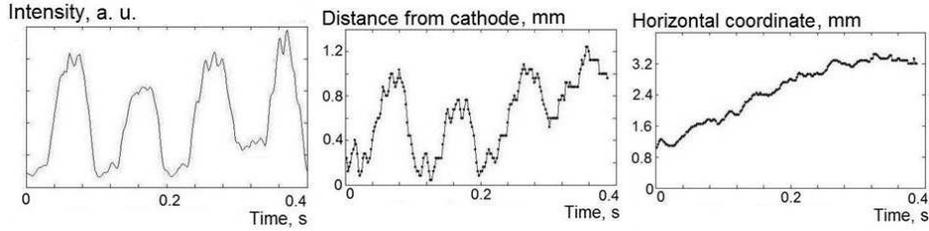}%
 \caption{\label{singlepart} Comparison between temporal behavior of sheath luminescence intensity and results of tracking for single particle, when additional 10Hz sinusoidal bias is applied to the cathode. Distance from cathode corresponds to vertical coordinate (movie online).}%
 \end{figure}

In the folowing, periodic sheath expansions and contractions were deliberately induced by adding a low amplitude (1W) oscillating component to the 10W power applied to RF generator. Power oscillations cause oscillations of plasma luminescence intensity as well. When power increases, luminescence intensity increases as well and the same does thickness of the sheath. Under our plasma conditions, sheath thickness linearly fits the cathode bias voltage, as it is shown in Fig.~\ref{figsheath}. Therefore, measuring temporal behavior of luminescence intensity directly from recorded movies one actually monitors temporal behavior of plasma sheath itself.
In order to enable massive dust grains to follow these fluctuations, it is necessary to apply low frequency oscillations, i.e. with a lower frequency than the characteristic dynamic scale of dust. Dust plasma frequency is about 100Hz for 15$\mu$m grains and 15Hz for 100$\mu$m grains. In the following, 10Hz power oscillations were applied. Visual inspection of movies recorded in these conditions reveals that when sheath region expands, dust particles move away from cathode and vice versa. This evolution is illustrated by Fig.~\ref{singlepart}, which represents the temporal evolution of sheath luminescence intensity as well as the temporal evolution of the coordinates of one single particle. While correlation between particle vertical displacement and sheath luminescence is obvious (in that case, correlation is 80\%), the horizontal displacement of particle is almost unaffected by sheath oscillation. 

In order to statistically estimate the global reliability of our detection and tracking procedure, we propose to compute the averaged cross-correlation $C_{l,y}$ between sheath luminescence variations and the vertical component of tracked particles:
  $$C_{l,y} = \frac{\sum_{t}^{T-1}[(l(t)-\overline{l})*(y(t+\tau)- \overline{y})]}{ \sqrt{ \sum_{t}(l(t)-\overline{l})^{2}} * \sqrt{ \sum_{t}(y(t+\tau)-\overline{y})^{2}} } $$
where $l(t)$ is the time series of light fluctuations spatially averaged in the region of interest, $y(t)$ is the vertical coordinate of a single particle, $\overline{l}$ and $\overline{y}$ are the time averaged values of these time series, and $T$ the length of the time series. In order to avoid spurious effect due to signals periodicity, delay $\tau$ is taken equal to 0 in the following. Only time series longer than one oscillation period are taken into account.
With the normalization used $C_{l,y}$ is bounded between $-1$ and 1. $C_{l,y}=1$ would mean that light and particle oscillation are perfectly correlated without any phase shift (i.e. $l(t)=y(t)$), and $C_{l,y}=-1$ that they are perfectly anticorrelated (i.e. $l(t)=-y(t)$), $C_{l,y}=0$ indicating the total absence of correlation.

Thus, if particles are correctly tracked, the averaged correlation $\langle C_{l,y} \rangle$ will be close to 1, whereas misdetection and mistracking will lower the averaged correlation value.
%
%
\subsection{Statistical validation based on dedicated experiments} 
\label{expvalidation}
\begin{figure}
      \includegraphics[width=8cm]{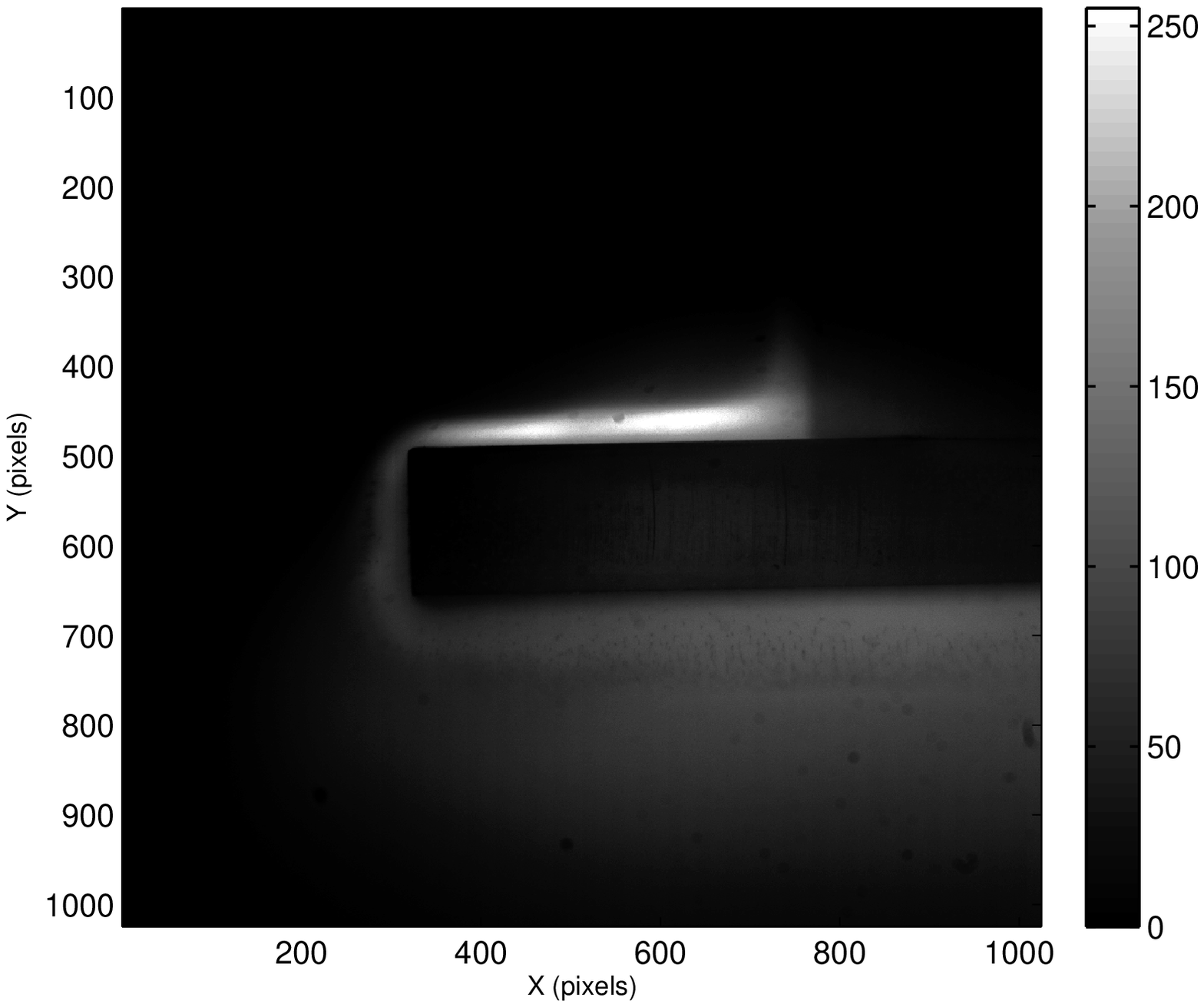}
\hfill
      \includegraphics[width=8cm]{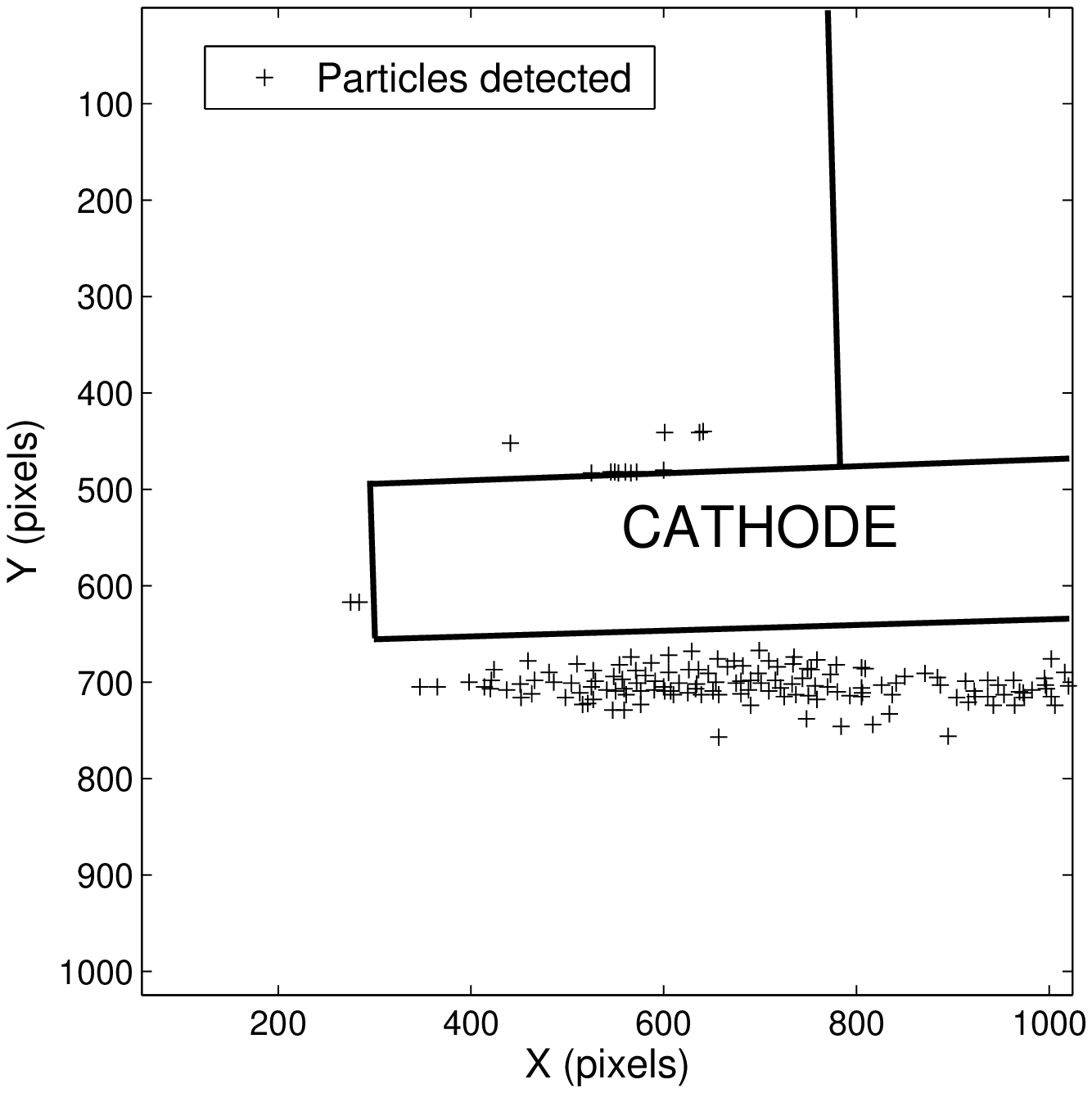}
\caption{Example of automatic detection using local adaptive thresholding: original camera frame (left) and automatically detected particles (right). Only few misdections occur at the upper edge of the cathode.}\label{contour}
 \end{figure}

For validation purpose, the analysis presented in this section focuses on the ROI depicted in figure~\ref{ROI}, i.e. a region of the sheath where particles exhibit clear vertical oscillations due to vertical electric field. In this ROI there is no sharp luminosity gradient due to, e.g. the edge of the cathode, but plasma luminosity is inhomogeneous in the vertical direction and its value strongly varies in time. Given these characteristics and the small size of the ROI, global thresholding applied to this ROI yields rather similar results as locally adaptive thresholding. In the rest of this section, $T_{hr,auto}$ refers to automatic threshold computed using global threshold applied to the ROI. If the entire frame is considered, locally adaptive threshold has to be chosen instead, as illustrated by Fig.\ref{contour}.

\begin{figure}
 \includegraphics[width=0.5\textwidth]{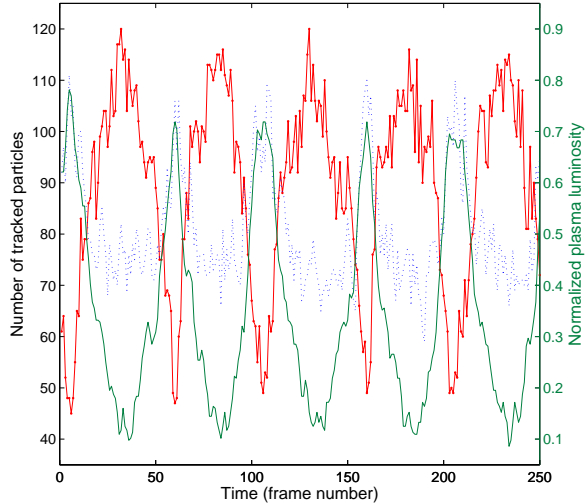}%
 \caption{\label{npart} Temporal variation of the number of tracked particles (left axis) with fixed normalized threshold value $T_{hr}=0.0025$ (red firm line with dots) and with automatic threshold (blue dotted line), compared with the temporal evolution of plasma normalized luminosity (right axis, green firm line). Before processing, background is removed and spatial filter is applied.}%
 \end{figure}

\begin{figure}
 \includegraphics{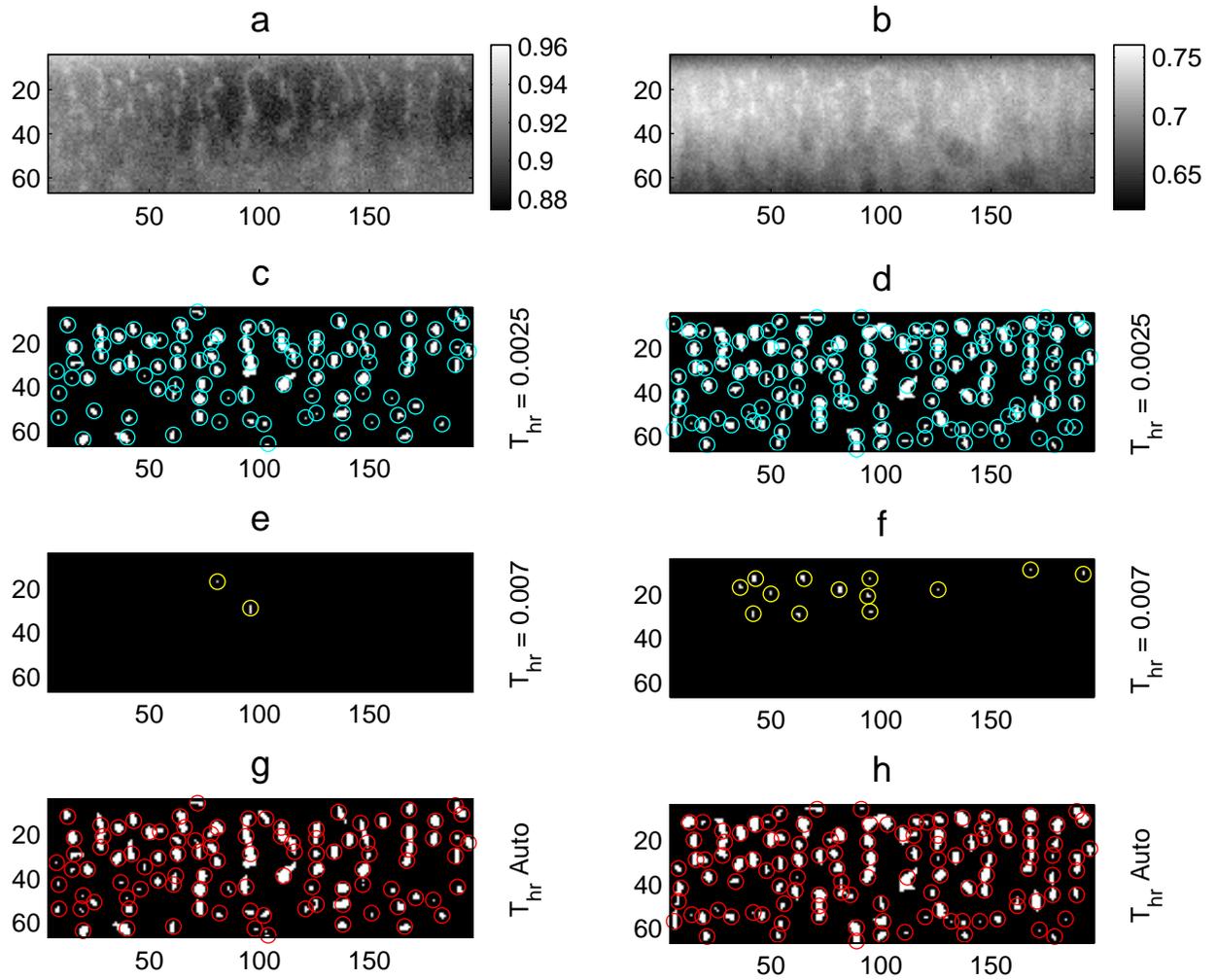}%
 \caption{\label{frame5} Particle detection in a small window beneath cathode: 5$^{th}$ frame of the movie analyzed in Fig.\ref{npart}. (a) raw image, (b) the same image after background removal (axes are in pixel units). Greyscale represents the normalized light intensity. Left side: result of image segmentation with fixed normalized threshold $T_{hr}=0.0025$ (c), with $T_{hr}=0.007$ (e) , and with automatic threshold (g). Right side: result of image segmentation applied to image (b) with the same thresholds $T_{hr}=0.0025$ (d), $T_{hr}=0.007$ (f) , and automatic (h).}%
 \end{figure}

\begin{figure}
 \includegraphics{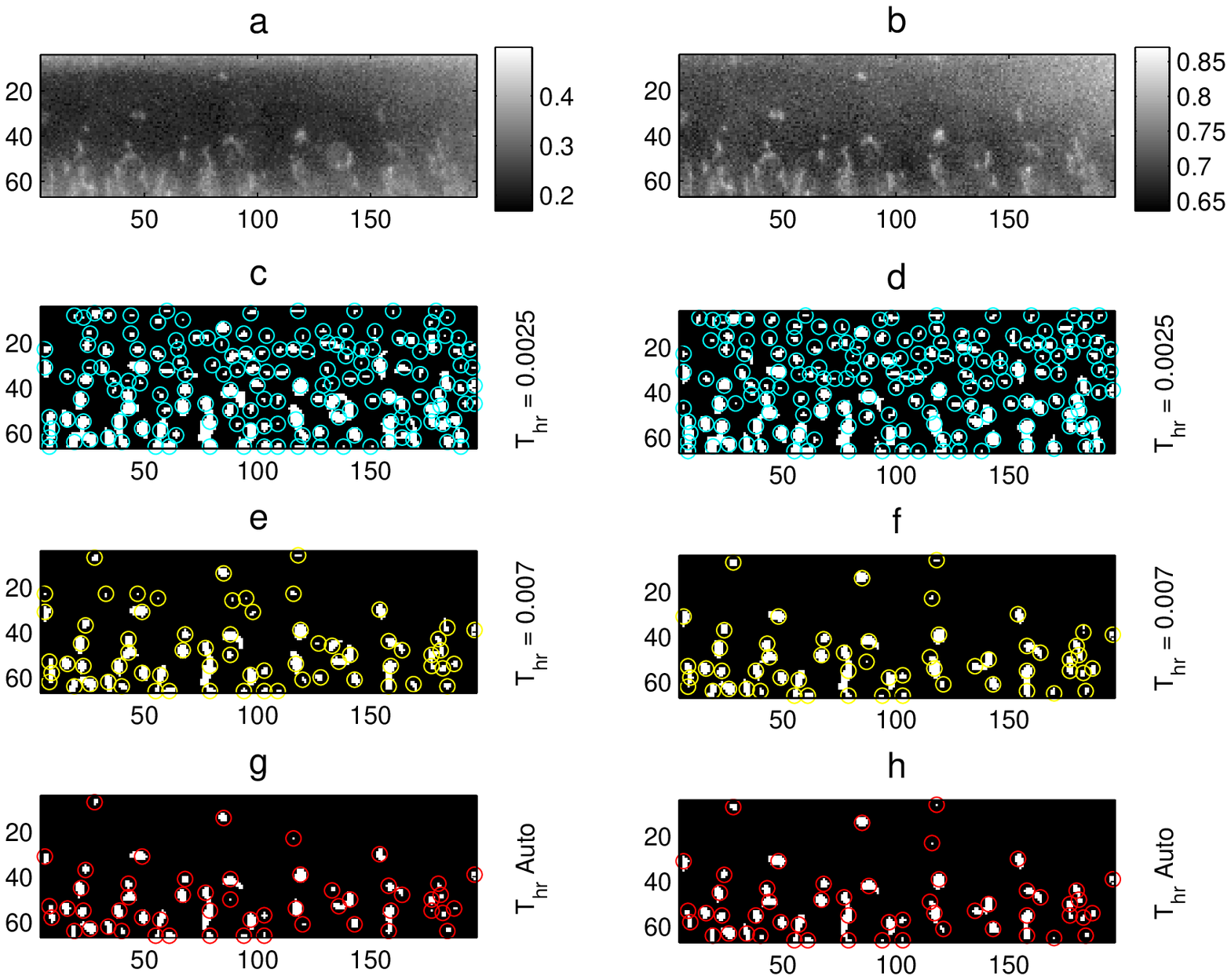}%
 \caption{\label{frame35} Same graphs as Fig.\ref{frame5}, but applied to 35$^{th}$ frame of the movie analyzed in Fig.\ref{npart}.}%
 \end{figure}

In order to assess the efficiency of particle detection under variable illumination conditions, the temporal evolution of background luminosity is confronted to the temporal evolution of the total number of tracked particles for two different thresholds. Background luminosity is spatially averaged in the ROI and normalized to 1. In order to apply fixed threshold, it is necessary to preliminary process camera frames in order to highlight moving particles: background is removed and spatial bandpass filter is applied (filter parameters $L_{noise}=1$ and $L_{Max}=4$). Fixed value $T_{hr}=0.0025$ (corresponding to light intensity of 0.12 in the unfiltered frame) is manually chosen after visual inspection of the movie. As shown in Fig.\ref{npart}, particle detection depends very much on the variations of background illumination. When plasma illumination is strong, the number of tracked object is minimum with fixed threshold and maximum with automatic threshold, whereas the opposite is seen when plasma illumination is low. In order to understand this result, both situations are analyzed more in details in Figs~\ref{frame5} and \ref{frame35}. As seen in Fig.\ref{frame5}(a), when plasma illumination is high the dynamic of the image is rather low: maximal pixel intensity is only 10\% higher than the minimal one, resulting in low contrasted particles. In Fig.\ref{frame35}(a), where plasma illumination is low, difference between lowest and highest intensity pixels is about 30\%, so that particles are much more contrasted. As seen from Figs.\ref{frame5}(b) and \ref{frame35}(b), removing background lowers the average intensity of luminous frames and enhances the average intensity of dark frames (when image background is removed, the pixel intensity is shifted in order to avoid negative values). This observation explains the variations noted in Fig.\ref{npart} when $T_{hr}$ is fixed. Since automatic threshold explicitely varies according to the average image intensity, it is almost unaffected neither by background illumination nor by background removal. The temporal variations observed in Fig.\ref{npart} are actually due to particles entering and leaving the ROI, as can be seen from Figs.\ref{frame5} and \ref{frame35}. Another important point illustred by these 2 figures is that automatic threshold is much less influenced by noise than fixed threshold. As can be seen from Figs.\ref{frame5}(g-h) and \ref{frame35}(g-h), most detections obtained with $T_{hr,auto}$ indeed correspond to true particles, whereas a large fraction of events detected with $T_{hr}=0.0025$ correspond to noisy pixels. When a more selective value $T_{hr}=0.007$ is taken, many dust particles are not detected during strong illumination phase, whereas misdetection remain in the low illumination phase (cf. Figs.\ref{frame5}(e-f) and \ref{frame35}(e-f)). As a comparison, the extremal values taken by $T_{hr,auto}$ for this analysis were $[0.0012-0.0062]$.

\begin{figure}
      \includegraphics[width=8cm]{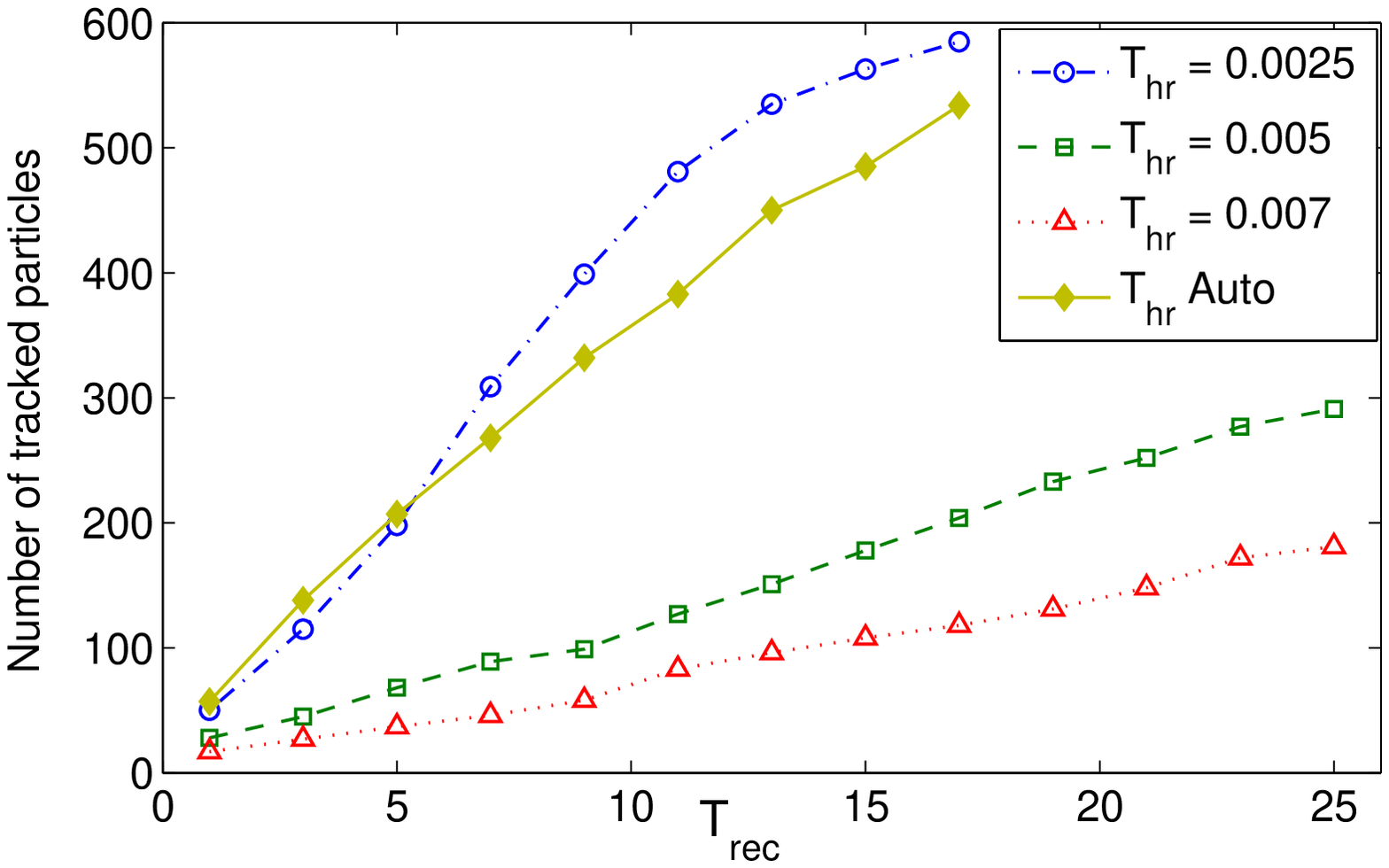}
 \caption{\label{nbtrec} Number of particles tracked for more than 50 frames function of $T_{rec}$ for three fixed threshold values (triangle, square, and circle) and automatic threshold (losange).}
 \end{figure}

As it is mentionned in Sec. III.C.2, one very important tracking parameter is the maximum recovery time $T_{rec}$. In order to determine the acceptable range of values of this parameter, the same movie was analyzed for different values of $T_{rec}$, and trajectories longer than one oscillation period, i.e. more than 50 frames, were analyzed quantitatively and qualitatively. The total number of such trajectories, function of $T_{rec}$ is plotted for different thresholds in Fig.\ref{nbtrec}. On the one hand, whereas more particle events are found by using lowest fixed value of $T_{hr}$ compared to automatic threshold (Figs.~\ref{npart}-\ref{frame35}), the number of trajectories longer than 50 frames is very similar. This is due to the impossibility for the tracking algorithm to link successive positions of spurious detections caused by inadequate value of $T_{hr}$. On the other hand, when higher fixed value of $T_{hr}$ is chosen, the number of long trajectories is drastically lower, in accordance with previous observation that high values of $T_{hr}$ are too selective. Whatever $T_{hr}$ in this range of values of $T_{rec}$, the longer time particle can be lost, the higher the number of trajectories longer than 50 frames (when $T_{rec}$ is larger than 25 frames, this number decreases due to the recombination of long trajectories). 

\begin{figure}
     \includegraphics[width=8cm]{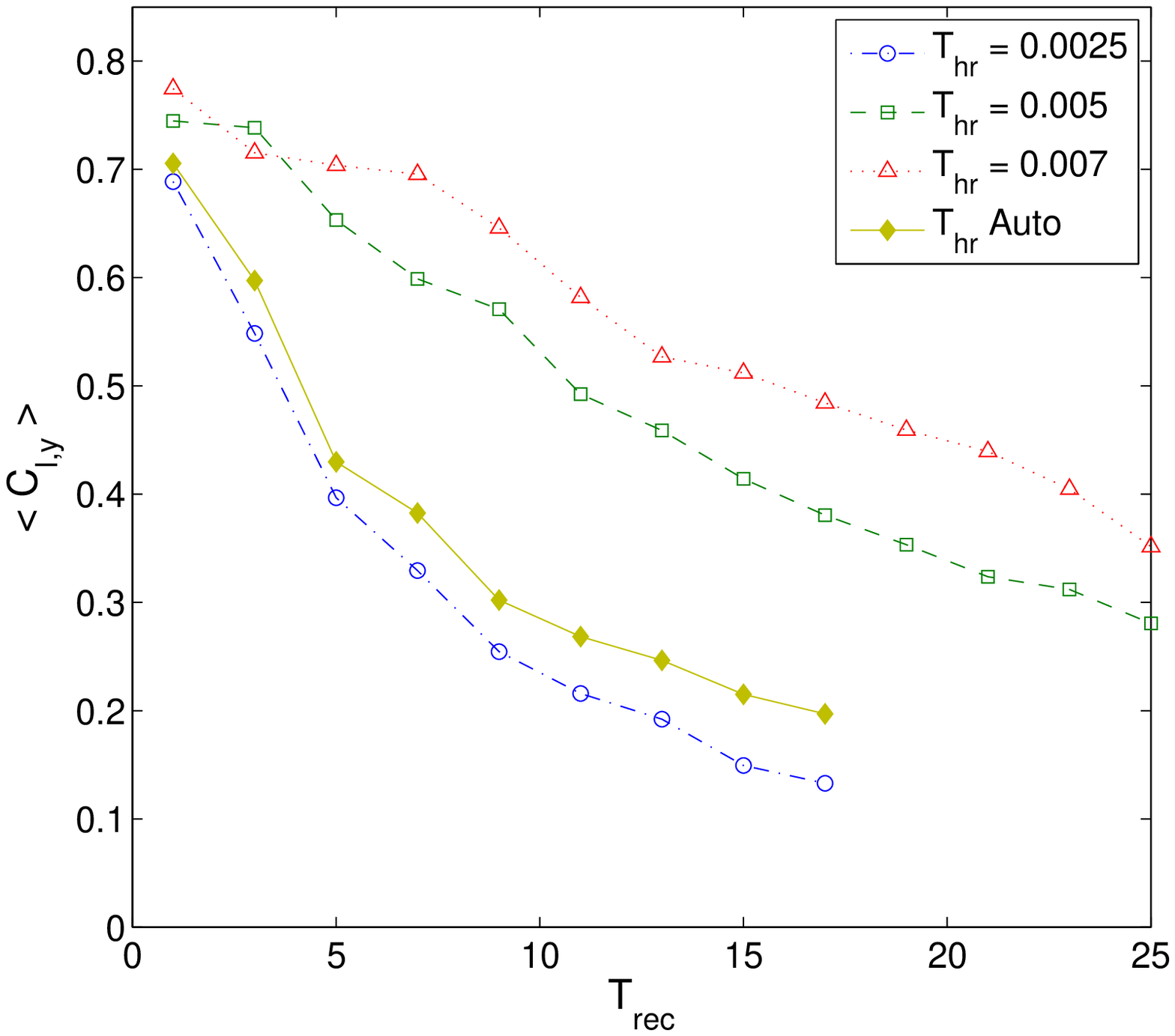}
\caption{\label{cortrec} Average correlation $<C_{l,y}>$ function of $T_{rec}$, for the same cases as Fig.~\ref{nbtrec}.}
 \end{figure}

\begin{figure}
 \includegraphics{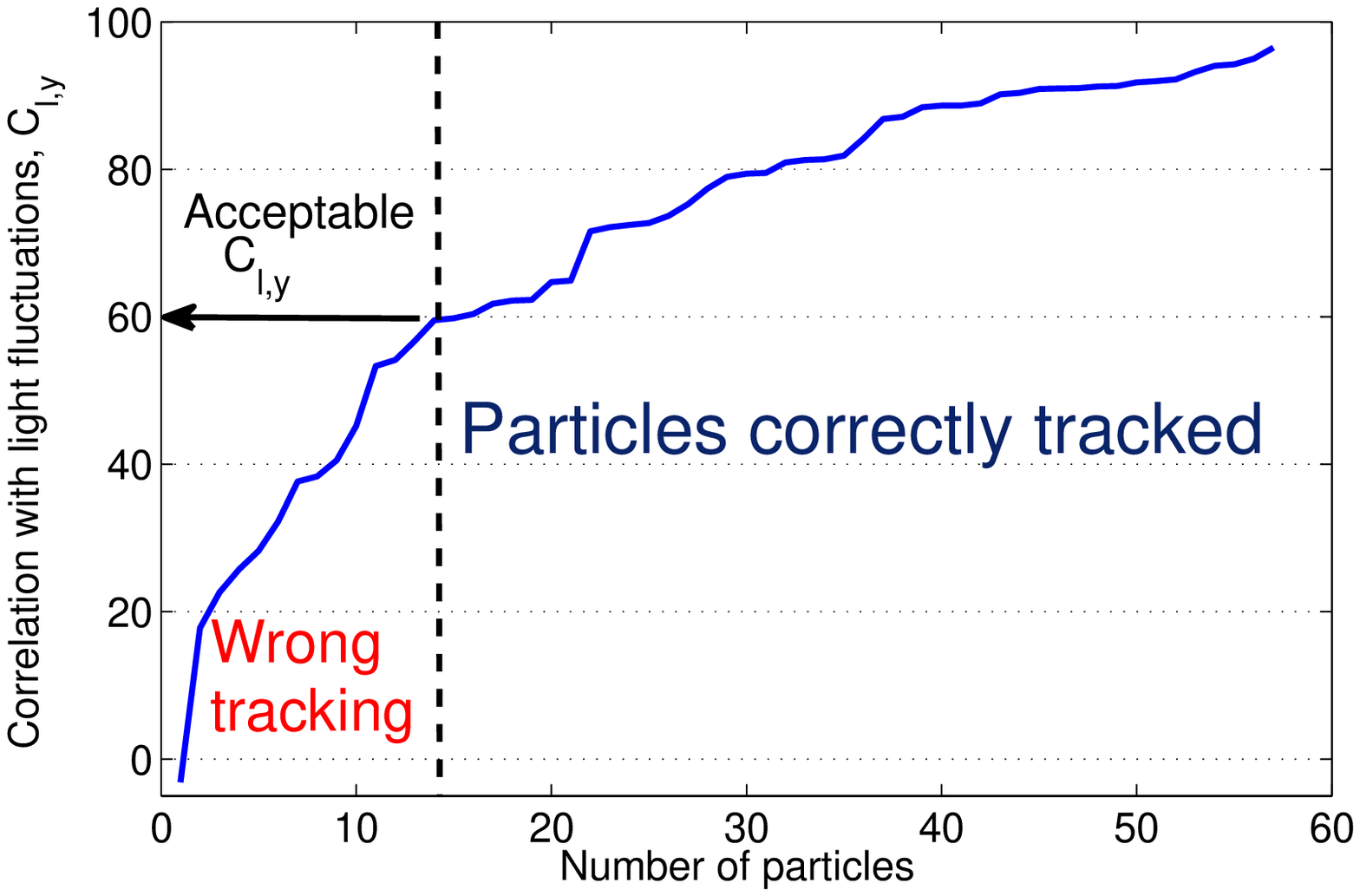}%
 \caption{\label{goodcorr} Correlation $C_{l,y}$ (in \%) for particles tracked more than 50 frames with automatic threshold ($T_{rec}=3$). Individual manual checking indicates that particles with $C_{l,y} \geq$ 60\% are correctly tracked. On the opposite, most of particles with $C_{l,y} \leq$ 60\% have uncorrect trajectories.}%
 \end{figure}

In order to estimate the reliability of these trajectories, $C_{l,y}$ is computed for each particle $p$. For each set of $T_{rec}$, the average value  $<C_{l,y}> = \frac{1}{N}\sum_{p=1}^{N}C_{l,y}(p)$, where $N$ is the total number of particles tracked more than 50 frames, is plotted in Fig.\ref{cortrec}. Obviously, $<C_{l,y}>$ rapidly decays when $T_{rec}$ is increased, and the more selective $T_{hr}$ is, the higher $<C_{l,y}>$. There are two possible ways to analyze these results. $<C_{l,y}>$ may decay because trajectory segments are not correctly linked when $T_{rec}$ is large, causing confusion between several particles. Another possibility is that $<C_{l,y}>$ decays because more particles are taken into account, which do not clearly follow sheath fluctuations (e.g. massive dust particles). It is thus necessary to determine the range of acceptable values of $<C_{l,y}>$. For that purpose, probability distribution functions (PDF) of $<C_{l,y}>$ have been analyzed. More precisely, trajectories belonging to different parts of the PDFs have been manually checked. The result obtained for the PDF shown in Fig.~\ref{goodcorr} can be generalized to other PDFs. It shows that particle trajectories with a correlation $<C_{l,y}>$ larger than 60\% are correct, whereas most of particles with a lower correlation are not. In this case, more than $2/3$ of particles are correctly tracked.
 
According to this result, a new look at Figs.\ref{nbtrec} and \ref{cortrec} allows concluding that best performance is obtained by using $T_{rec} = 3$ with automatic threshold, since this setting yield the largest number of correct trajectories. While the difference with results obtained with $T_{rec} = 3$ and fixed threshold $T_{hr} = 0.0025$ might seem rather poor, it should be kept in mind that only trajectories longer than 50 frames were analyzed. One very important benefit of automatic threshold, which does not appear in these figures, is the very low ratio of false detections. As a result, using $T_{hr,auto}$ does not only provide more accurate results, but also a significant gain of time and ressources. Whereas the analysis of a 1000 frames movie with a pentium IV computer requires about 10s with $T_{hr,auto}$, the time needed with $T_{hr} = 0.0025$ exceeds several minutes, due to the largest number of possible recombinations during the tracking procedure.

%
%
\subsection{Statistical validation based on simulations}
In order to assess the efficiency of the tracking procedure independently from the detection procedure, virtual dust particles with known trajectories were simulated in a finite size box. Perfectly known trajectories were confronted to the ones obtained with the tracking procedure applied to the array containing the list of mixed particle positions. Tsallis distribution function is chosen since many recent studies have shown that dust particles in very similar physical systems exhibit superdiffusive behaviour characterized by such a velocity distribution \cite{liu2, nuno, raty}. Results for 1000 independent realizations and for two different sets of velocity distributions are adducted in Table~\ref{tablesimu}. For a pure Tsallis velocity distribution, 30\% of particles are perfectly tracked over the 500 frames of the simulations. On average, particles are correctly tracked during 75\% of their trajectory, i.e. 375 consecutive frames. Mistracking is caused by recombination of different particle trajectories due to occlusions (i.e. confusion between particles distant of less than $D_{M}$), or to particle jumping more than the expected maximal displacement ($D_{M} = 5$). When oscillations are forced in one given direction, half of trajectories are perfectly extracted and the global accuracy raises up to about 85\%. This is mainly due to the fact that collective motion in at least one direction significantly reduces the probability of particle occlusion. These simulation results are very close to results presented in section~\ref{expvalidation} for oscillating dust particles in laboratory experiment, demonstrating the excellent performance of the detection procedure. They also highlight that the tracking efficiency is very sensitive to particle dynamics. 

\begin{table}
\begin{tabular}{|l|c|c|c|c|}
\hline
Velocity & Ratio of perfect & Global accuracy & Number of & Number of\\
distribution & trajectories & & occlusions & particle jumps\\
\hline
Tsallis(x,y) & 30\% & 75\% & 608 & 127\\
\hline
Tsallis(x,y) & 50\% & 85\% & 56 & 57\\
+ y oscillations & & & &\\
\hline
\end{tabular} 
\caption{\label{tablesimu} Comparison between known and reconstructed trajectories obtained with the tracking procedure. 60 particles in a 60$\times$60 px$^{2}$ box were considered for 500 frames.}
\end{table}

Obviously, improving the overall performance of the detection and tracking algorithm requires to focus more particularly on the improvement of the tracking procedure. This has to be done by extrapolating individual trajectories in order to minimize confusion due to particle occlusions. Extrapolation can be done either during the tracking procedure \cite{ouel} or after complete execution of the algorithm. First solution probably yields more accurate results but its main drawback is that it may require excessive computational resources when particle density is large, since it becomes necessary to store previous positions of all the particles for all possible combinations. Second solution consists in using extrapolation \textit{a posteriori} to correctly recombine trajectory segments. This solution requires low computational resources, but in order to lower the amount of uncorrect segments, $T_{rec}$ should be kept very low (typically $0 \leq T_{rec} \leq 2$). In both cases, the function which has to be minimized during the extrapolation has to be carefully chosen according to the proper dynamic of the particulate medium. In simple cases linear extrapolation can be good enough \cite{ends}. More sophisticated solutions have demonstrated good efficiency for passive tracers in turbulent flows \cite{ouel}, but to the best of our knowledge new solutions have to be found for very stochastic systems like the one investigated in this paper.

%
%
\section{Discussion and Summary}
An algorithm has been written, which makes it possible to automatically detect and track dust particles in plasmas, even under challenging observation conditions. An experimental benchmark has also been proposed to overcome the difficult issue of code validation. Although simulations remain the best solution to test the efficiency of different tracking procedures, the main advantage of the proposed experiment is that it enables evaluating the complete detection and tracking procedure, in real experimental conditions. Hence, reference data can be used to easily monitor the impact of modifications to be brought to the code. Presented results demonstrate that rather simple advanced thresholding techniques show a good capability to discriminate particles from complex images with fluctuating background and noise, while keeping the requirement for computational resources very low. In addition, adaptive threshold makes it useless to remove background, thus allowing detection of slow or stationary small objects as well. This property is currently being used to analyze not only particle transport but also hot spots in fusion plasmas \cite{ends}, demonstrating the ability of our code to run in very different conditions. Nevertheless, some improvements of the tracking procedure are necessary in order to obtain more reliable trajectories in dense particulate media with stochastic motion. Taking into account stereoscopic data from several cameras is another ongoing task, which should contribute to such an improvement by providing 3D trajectories.

The algorithm, coded in Matlab programming language, is available from the authors on request.

\begin{acknowledgments}
This work, supported by the European Community under the contract of Association
between EURATOM, CEA, and the French Research Federation for fusion studies, was
carried out within the framework of the European Fusion Development Agreement. The views and opinions expressed herein do not necessarily reflect those of the European Commission. Financial support was also received from the French National Research Agency through contract ANR-08-JCJC-0068-01, and from a Bonus Qualité Recherche (BQR) of the Henri Poincaré University.
\end{acknowledgments}

\bibliographystyle{aipnum4-1}
\bibliography{paper}

\end{document}